# Large spin splitting and piezoelectricity in a two-dimensional topological insulator Al$_2$SbBi with double-layer honeycomb structure

D. Q. Fang[1*], H. Zhang[1], D. W. Wang[2,3]

[1]MOE Key Laboratory for Nonequilibrium Synthesis and Modulation of Condensed Matter, School of Physics, Xi'an Jiaotong University, Xi'an 710049, China

[2]School of Microelectronics & State Key Laboratory for Mechanical Behavior of Materials, Xi'an Jiaotong University, Xi'an 710049, China

[3]Key Lab of Micro-Nano Electronics and System Integration of Xi'an City, Xi'an Jiaotong University, Xi'an 710049, China

Abstract

Two-dimensional materials provide remarkable platforms to uncover intriguing quantum phenomena and develop nanoscale devices of versatile applications. Recently, AlSb in the double-layer honeycomb (DLHC) structure was successfully synthesized exhibiting a semiconducting nature [ACS Nano 15, 8184 (2021)], which corroborates the preceding theoretical predictions and stimulates the exploration of new robust DLHC materials. In this work, we propose a Janus DLHC monolayer Al$_2$SbBi, the dynamical, thermal, and mechanical stabilities of which are confirmed by first-principles calculations. Monolayer Al$_2$SbBi is found to be a nontrivial topological insulator with a gap of about 0.2 eV, which presents large spin splitting and peculiar spin texture in the valence bands. Furthermore, due to the absence of inversion symmetry, monolayer Al$_2$SbBi exhibits piezoelectricity and the piezoelectric strain coefficients $d_{11}$ and $d_{31}$ are calculated to be 7.97 pm/V and 0.33 pm/V, respectively, which are comparable to and even larger than those of many piezoelectric materials. Our study suggests that monolayer $Al_2SbBi$ has potential applications in spintronic and piezoelectric devices.

*fangdqphy@xjtu.edu.cn

I. Introduction

Two-dimensional (2D) materials have attracted intense interest due to their novel physical properties, such as superconductivity in single-layer FeSe films [1], spin-valley coupling in layered transition metal dichalcogenides [2], quantum spin hall state in monolayer 1T'-WTe$_2$ [3], and Ising ferromagnetism in monolayer CrI$_3$ [4], offering promising candidates for applications in next generation electronic devices. The further discovery of 2D materials provides the potential of unraveling quantum phenomena and could lead to applications in multifunctional quantum devices.

Many high-quality 2D materials can be derived from layered three-dimensional (3D) materials by mechanical exfoliation [5]. Nevertheless, a recent experiment [6] showed that by using graphene capping layer one can grow GaN, a traditional wide-bandgap 3D semiconductor, into 2D form on a SiC substrate, providing the foundation to realize 2D forms for other semiconductors that are not traditionally 2D. The theoretical study by Lucking *et al*. suggested that in the ultrathin limit the majority of traditional III-V, II-VI, and I-VII semiconductors are stable in a 2D double-layer honeycomb (DLHC) structure [7]. These proposed 2D materials can exhibit exotic physical properties, including the topological transition of electronic structure and the formation of an excitonic insulator.

Recently, 2D AlSb in the DLHC structure has been successfully synthesized through molecular beam epitaxy growth on a graphene-covered SiC(0001) substrate, possessing a band gap of 0.93 eV revealed by the tunneling conductance spectrum [8], which verifies the feasibility of theoretically predicted DLHC materials. Furthermore, 2D

AlSb was predicted by using first-principles calculations to have an excitonic instability and sufficient carrier mobility [9]. Xia *et al*. proposed a new semimetal AlSb monolayer with a rectangular lattice that has multiple nodal-loops and extraordinary transport properties under uniaxial strain [10].

It is noted that Janus monolayer MoSSe has been synthesized experimentally by replacing the top-layer S atoms of $MoS_2$ monolayer with Se atoms [11]. Due to the breaking of out-of-plane mirror symmetry Janus transition metal dichalcogenide monolayers exhibit distinct properties, such as large Rashba spin splitting around the Γ point [12], out-of-plane piezoelectricity [13,14], and superior photocatalytic performance [15]. Inspired by this, we explore, in this work, the stability and properties of Janus DLHC monolayer $Al_2SbBi$ obtained by replacing the top-layer Sb atoms of DLHC monolayer AlSb with Bi atoms.

## II. Computational details

Our first-principles density functional theory calculations are performed using the Vienna *ab initio* simulation package (VASP) [16] code. The exchange-correlation potential is described within the generalized gradient approximation (GGA) parametrized by Perdew, Burke, and Ernzerhof (PBE) [17]. The electron-ion interaction is described by the projector augmented-wave method [18,19]. A vacuum region of 12 Å is employed to avoid the interactions between neighboring images. For structural relaxations, plane wave energy cutoff is set to 500 eV and Brillouin-zone (BZ) sampling is performed using a Γ-centered 12×12×1 **k**-point grid [20]. The electronic self-consistency convergence criterion is set to $10^{-6}$ eV and $10^{-8}$ eV for structural relaxation and static calculations, respectively. Dipole correction is exerted in the out-of-plane direction according to the method described in Ref. [21]. All atoms are relaxed until the Hellmann-Feynman forces are less than 0.005 eV/Å. The elastic stiffness tensor is obtained by performing six finite distortions of the lattice and deriving the elastic constants from the strain-stress relationship [22]. The piezoelectric stress tensor is calculated using the density functional perturbation theory (DFPT) and the PBE functional including spin-orbit coupling (SOC). To compensate for the band gap underestimation of the PBE functional, we perform additional Heyd-Scuseria-Ernzerhof (HSE06) screened hybrid functional [23] calculations for the electronic structure.

To examine the dynamical stability of monolayer $Al_2SbBi$, we carry out phonon calculations using the DFPT as implemented in the QUANTUM ESPRESSO code [24–26]. Kinetic energy cutoff for wavefunctions is set to 80 Ry and standard solid-state pseudopotentials, i.e., SSSP[v1.2.1][Precision][PBE]) [27], are used. The vibrational Brillouin zone is sampled using 8×8×1 **q**-point grid.

### III. Results and Discussion

A. Structural property and stability analysis

The structure of Janus DLHC monolayer $Al_2SbBi$ is shown in Figs. 1(a) and 1(b). This structure is composed of AlSb and AlBi single-layers bound together through the interlayer bonds, having a space group P3m1 (No. 156) without inversion symmetry.

The optimized in-plane lattice parameters are predicted as $a = b = 4.36$ Å. The lengths of Al-Bi bonds in the basal plane and out-of-plane direction are 2.796 Å and 2.945 Å, respectively, slightly larger than those of Al-Sb bonds, i.e., 2.748 Å and 2.831 Å. Figure 1 (c) depicts the electron localization function (ELF) of monolayer Al$_2$SbBi, showing more electron localization around the Sb atoms. This is also reflected in the Bader charge analysis, i.e., the total amount of charge transfer from the surrounding Al atoms to the Sb (1.38e) is larger than that to the Bi (1.11e).

To evaluate the stability of this structure, we first compute the cohesive energies of DLHC monolayer Al$_2$SbBi and AlSb by using the following formula

$$E_{coh} = (xE_{Al} + yE_{Sb} + zE_{Bi} - E_{tot})/(x + y + z), \qquad (1)$$

where $E_{tot}$, $E_{Al}$, $E_{Sb}$, and $E_{Bi}$ are the energies of unit cell and isolated Al, Sb, and Bi atoms including the spin-orbit interaction, respectively, and x, y, and z are the number of Al, Sb, and Bi atoms in unit cell. The cohesive energies of monolayer Al$_2$SbBi and AlSb are 2.85 eV and 3.08 eV, respectively. For comparison, using the same computational method, the cohesive energies of DLHC monolayer CuI and AgI are computed to be 2.43 eV and 2.05 eV, respectively, which are smaller than that of monolayer Al$_2$SbBi. Monolayer AlSb, CuI, and AgI have been successfully synthesized recently [8,28]. Thus, from the point of view of cohesive energy, it is likely to synthesize monolayer Al$_2$SbBi when restricted to 2D.

Figure 2(a) shows the calculated phonon dispersion along the high-symmetry lines for monolayer Al$_2$SbBi. There is no imaginary frequency in the phonon spectrum, indicating the dynamical stability of this material. To check the thermal stability of monolayer Al$_2$SbBi, we perform molecular dynamics (MD) simulations at 300 K and 500 K in the canonical ensemble. A 4×4 supercell is used in the MD simulation. Figure 2(c) shows small energy fluctuation during the simulation at 300 K. The frameworks of monolayer Al$_2$SbBi at the end of the simulations at 300 K [Fig. 2(d)] and 500 K [Supplementary Fig. S2(b)] are well-kept, confirming the thermal stability at ambient conditions.

We verify the mechanical stability of monolayer Al$_2$SbBi by checking the elastic constants. Table I shows the calculated clamped- and relaxed-ion elastic stiffness coefficients $C_{11}$ and $C_{12}$. For monolayer Al$_2$SbBi, ionic relaxation makes the value of $C_{11}$ reduced and $C_{12}$ slightly increased compared to the clamped-ion case, which agrees with the trend for monolayer h-BN [29]. The results satisfy the necessary and sufficient elastic stability conditions for the hexagonal lattice [30]: $C_{11} > 0$ and $C_{11} > |C_{12}|$, confirming the mechanical stability of monolayer Al$_2$SbBi. The isotropic Young's modulus of monolayer Al$_2$SbBi is computed to be 48.9 N/m based on the relaxed-ion elastic coefficients [Supplementary Fig. S1(a)], much smaller than that of graphene (339 N/m) and h-BN (277 N/m) [31], indicating excellent flexibility. The isotropic Poisson's ratio is computed to be 0.37 [Supplementary Fig. S1(b)].

B. Electronic structure

The electronic band structure of monolayer Al$_2$SbBi is studied by comparing the PBE functional and HSE06 hybrid functional. As shown in Figs. 3(a) and 3(c), in the absence of SOC a metallic band structure is found using the PBE while a semiconducting behavior with a direct band gap of 0.066 eV at the Γ point occurs in the case of HSE06.

Once SOC is accounted for, both PBE and HSE06 show an insulating phase [Figs. 3(b) and 3(d)]. PBE gives an indirect band gap of 0.093 eV with the valence band maximum (VBM) located near the Γ point and the conduction band minimum (CBM) at the M point, while HSE06 presents a band gap of 0.195 eV with both the VBM and CBM located near the Γ point.

To elucidate the origin of the change of band structure, we analyze the orbitals' evolution at the Γ point. Figures 4(a)-(d) show the orbital decomposed HSE06 band structures. Without SOC, the CBM is mainly contributed by the Al s orbital and the VBM is dominated by the Bi $p_x$ and $p_y$ orbitals [Figs. 4(a) and 4 (b)], revealed also by the projected density of states in supplementary Fig. S3. With SOC, the degenerate $|p_{x,y}\rangle$ states are split into the $|p_{x+iy,\uparrow}, p_{x-iy,\downarrow}\rangle$ and $|p_{x-iy,\uparrow}, p_{x+iy,\downarrow}\rangle$ states between which the $|s\rangle$ state lower than the Fermi level is sandwiched [see Figs. 4(c), 4(d) and 4(e)]. Apparently, a band inversion occurs in the monolayer Al$_2$SbBi due to SOC.

To verify the nontrivial topological nature, we determine the topological invariant $\mathbb{Z}_2$ by calculating the hybrid wannier charge center (WCC) with the WannierTools [32] code. The tight-binding model is constructed using the Wannier90 [33] code based on the maximally-localized wannier functions and the HSE06 functional is employed. As shown in Fig. 5(a), WCC is crossed by any horizontal reference lines an odd number of times, indicating $\mathbb{Z}_2 = 1$. Thus, the system possesses the nontrivial topology and quantum spin Hall effect. We expect the presence of helical edge states that cross the bulk gap. We use the edge Green's function technique implemented in the WannierTools to calculate the edge spectral density with a semi-infinite system. In Fig. 5(b) we show the edge spectral density for the armchair edge of monolayer $Al_2SbBi$. Helical edge states inside the bulk gap, connecting the conduction bands with the valence bands, are clearly visible, corroborating the nontrivial topological phase.

Applying strain is one of the effective avenues in experiments to tune the electronic structure of 2D materials. Here we examine the biaxial strain effect on the magnitude of the band inversion at the Γ point ($E_{inv}$) and the band gap ($E_{gap}$) of monolayer $Al_2SbBi$, as shown in Fig. 6(a). The biaxial strain ε is defined as $(a - a_0)/a_0$, where $a_0$ and $a$ are the in-plane lattice parameters with unstrained and strained states, respectively. The positive strain indicates tension, while the negative strain denotes compression. When ε changes from 0 to -0.03, the value of $E_{inv}$ decreases almost monotonically from 0.422 eV to 0 eV and the topological gap decreases gradually. For ε ≤ -0.03, the system turns into a trivial phase. When ε changes from 0 to 0.03, the value of $E_{inv}$ increases significantly but the topological gap changes very little and saturates at around 0.2 eV. The band structures for ε = -0.03 and ε = 0.03 are shown in Figs. 6(b) and 6(c), respectively. To clarify the transition from the nontrivial state to trivial state, we compare the orbitals' evolution in the case of no strain with that under the compressive strain, as shown in Figs. 4(e) and 4(f). The compressive strain enlarges the energy difference between the $|s\rangle$ and $|p_{x,y}\rangle$ states. When the compressive strain reaches a sufficiently large amount, SOC cannot lead to a band inversion so that the trivial phase occurs.

Figure 7(a) presents the spin splitting at the top of valence bands with ΔE the spin splitting energy and $k_0$ the momentum offset around the Γ point. For monolayer $Al_2SbBi$, the spin splitting energy at the momentum offset of 0.079 Å$^{-1}$ is giant, as large as 0.131 eV predicted by the HSE06 calculation. The spin texture of the highest valence band around the Γ point is shown in Fig. 7(b). The spin direction is mainly in plane and the spin texture shows the clockwise and counterclockwise rotating spin directions at the central and outer region of Brillouin zone, respectively. This is similar to the Rashba spin splitting formed by two bands with the opposite in-plane helical spin texture [34]. We further study the biaxial strain effect on the spin splitting energy and momentum offset. As shown in Fig. 7 (c), the magnitude of spin splitting energy increases faster than that of momentum offset under the tensile strain. For ε = 0.03, the magnitude of ΔE (k$_0$) is 70% (42%) larger than that in the unstrained monolayer. The compressive strain makes both the splitting energy and momentum offset decreased. The large spin splitting, intriguing spin texture and strain tunability make monolayer Al$_2$SbBi a promising candidate in spintronic applications such as charge-to-spin conversion by the Rashba-Edelstein effect [35].

C. Piezoelectric properties

In non-centrosymmetric materials electric dipole moment can be induced by applied mechanical stress. Theoretical studies have demonstrated the enhanced piezoelectricity in many 2D materials [13,29,36]. Monolayer Al$_2$SbBi does not have an inversion center, so it is interesting to examine its piezoelectricity.

In the 2D limit, the z direction is strain/stress free and only the in-plane strain/stress is allowed. The relaxed-ion piezoelectric stress tensor $e_{ij}$ is described as the sum of the electronic and ionic contributions. The piezoelectric stress tensor $e_{ij}$ and piezoelectric strain tensor $d_{ij}$ are related through the elastic stiffness coefficient tensor $C_{ij}$ as follows [36]:

$$e_{ik} = d_{ij}C_{jk}. \tag{2}$$

For the DLHC structure, the piezoelectric strain tensor $d_{ij}$ is given by the relations [36]

$$d_{11} = \frac{e_{11}}{C_{11} - C_{12}}, \quad d_{31} = \frac{e_{31}}{C_{11} + C_{12}}. \tag{3}$$

As listed in Table II, the $e_{11}$ coefficient of monolayer $Al_2SbBi$ is calculated to be $2.838\times10^{-10}$ C/m, which is slightly smaller than that of monolayer 1H-MoS$_2$ [14] but considerably larger than that of recently predicted monolayer GaBiH [37] and SrAlGaSe$_4$ [38]. It is worth noting that monolayer GaBiH and SrAlGaSe$_4$ are predicted as piezoelectric quantum spin hall insulators (PQSHIs) with the coexistence of quantum spin hall states and piezoelectricity [37,38]. The existence of Sb and Bi atoms at each side makes the electric dipole moment in the out-of-plane direction nonzero. The $e_{31}$ is calculated to be $0.259\times10^{-10}$ C/m. The corresponding piezoelectric strain coefficients are calculated according to Eq. (3) using the relaxed-ion elastic stiffness coefficients and the $d_{11}$ and $d_{31}$ are found to be 7.97 pm/V and 0.33 pm/V, respectively. The magnitude of $d_{11}$ of monolayer $Al_2SbBi$ is about 1.9 times and 4.3 times larger than that of monolayer GaBiH and SrAlGaSe$_4$, respectively. Therefore, monolayer $Al_2SbBi$ is a potential candidate for PQSHI by applying uniaxial strain to achieve high-efficient electromechanical coupling in nanoscale devices such as sensors and actuators.

## IV. Conclusions

In this work, using first-principles calculations, we investigate the stability, electronic, and piezoelectric properties of Janus DLHC monolayer $Al_2SbBi$. Our results reveal that monolayer $Al_2SbBi$ possesses reasonable cohesive energy and dynamical, thermal, and mechanical stabilities, which is likely to be synthesized in experiment. Electronic band structure and wannier charge center calculations reveal that monolayer $Al_2SbBi$ is a nontrivial topological insulator with a gap of about 0.2 eV. The band inversion arises from the strong spin-orbit interaction in the Bi $p_{x,y}$ states. Moreover, large spin splitting and peculiar spin texture are found in the valence bands. Due to the absence of inversion symmetry, monolayer $Al_2SbBi$ exhibits piezoelectricity and the piezoelectric strain coefficients $d_{11}$ and $d_{31}$ are calculated to be 7.97 pm/V and 0.33 pm/V, respectively. Our work provides guidance for experimental synthesis efforts and future spintronic and piezoelectric applications for monolayer $Al_2SbBi$.

**Supplementary Material**

See supplementary material for Young's modulus, Poisson's ratio, molecular dynamics simulation at 500 K, projected density of states, and rectangular supercell for monolayer $Al_2SbBi$.

**Conflicts of interest**

There are no conflicts to declare.

**ACKNOWLEDGMENTS**

We acknowledge the financial support from the Natural Science Foundation of Shaanxi Province (Grant No. 2019JQ-240) and the National Natural Science Foundation of China (Grant Nos. 11604254, 11974268, and 12111530061). We also acknowledge the HPCC Platform of Xi'an Jiaotong University for providing the computing facilities.

Table I. Calculated clamped- and relaxed-ion elastic stiffness coefficients $C_{11}$ and $C_{12}$ (in units of N/m) of monolayer $Al_2SbBi$.

| Clamped ion | | Relaxed ion | |
|---|---|---|---|
| $C_{11}$ | $C_{12}$ | $C_{11}$ | $C_{12}$ |
| 71.0 | 19.7 | 56.9 | 21.3 |

Table II. Piezoelectric coefficients $e_{11}$, $e_{31}$ (in units of $10^{-10}$ C/m), $d_{11}$, and $d_{31}$ (in units of pm/V) for monolayer $Al_2SbBi$ studied in this work and some other 2D materials reported in the literature.

| | $e_{11}$ | $e_{31}$ | $d_{11}$ | $d_{31}$ |
|---|---|---|---|---|
| $Al_2SbBi$ (this work) | 2.838 | 0.259 | 7.97 | 0.33 |
| 1H-MoS$_2$ [14] | 3.7 | | 3.8 | |
| 1H-MoSTe [14] | 4.5 | 0.5 | 5.1 | 0.4 |
| GaBiH [37] | 1.071 | | 4.163 | |
| SrAlGaSe$_4$ [38] | 0.575 | 0.053 | 1.865 | 0.068 |

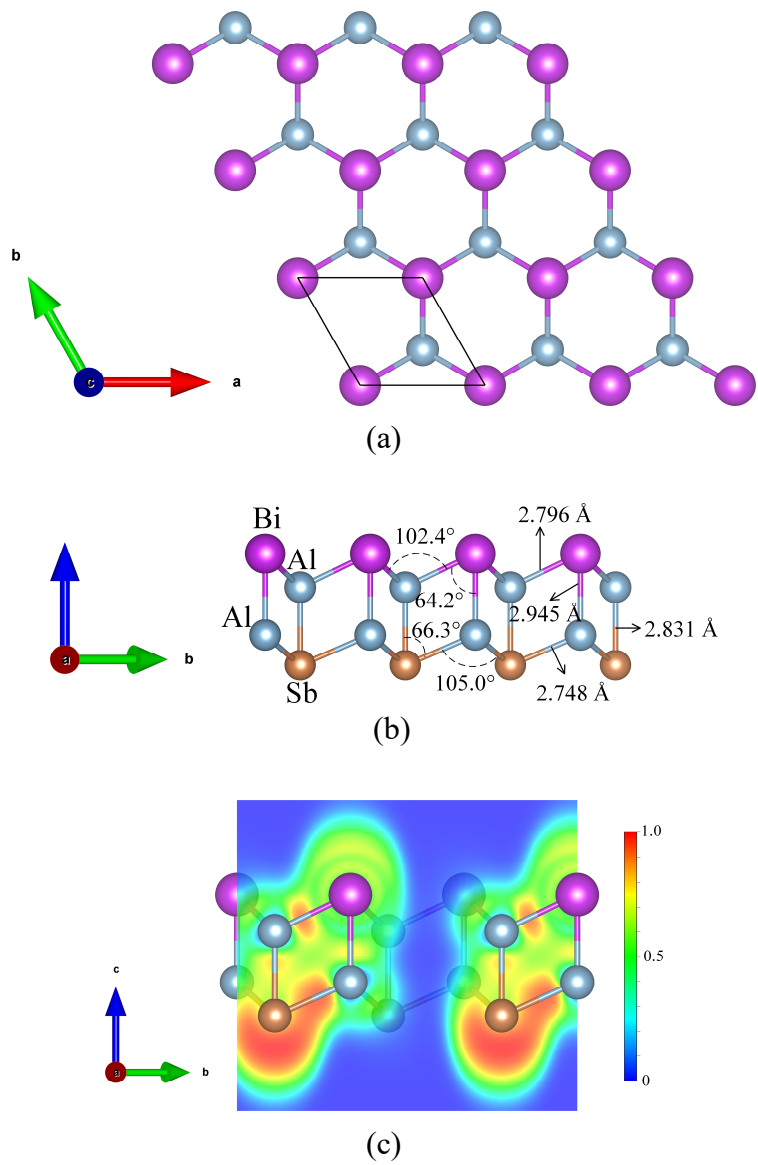

Fig. 1 Top view (a) and side view (b) of the atomic structure of monolayer $Al_2SbBi$ and the corresponding ELF (c). The black frame indicates the primitive unit cell, and the bond lengths and angles are given.

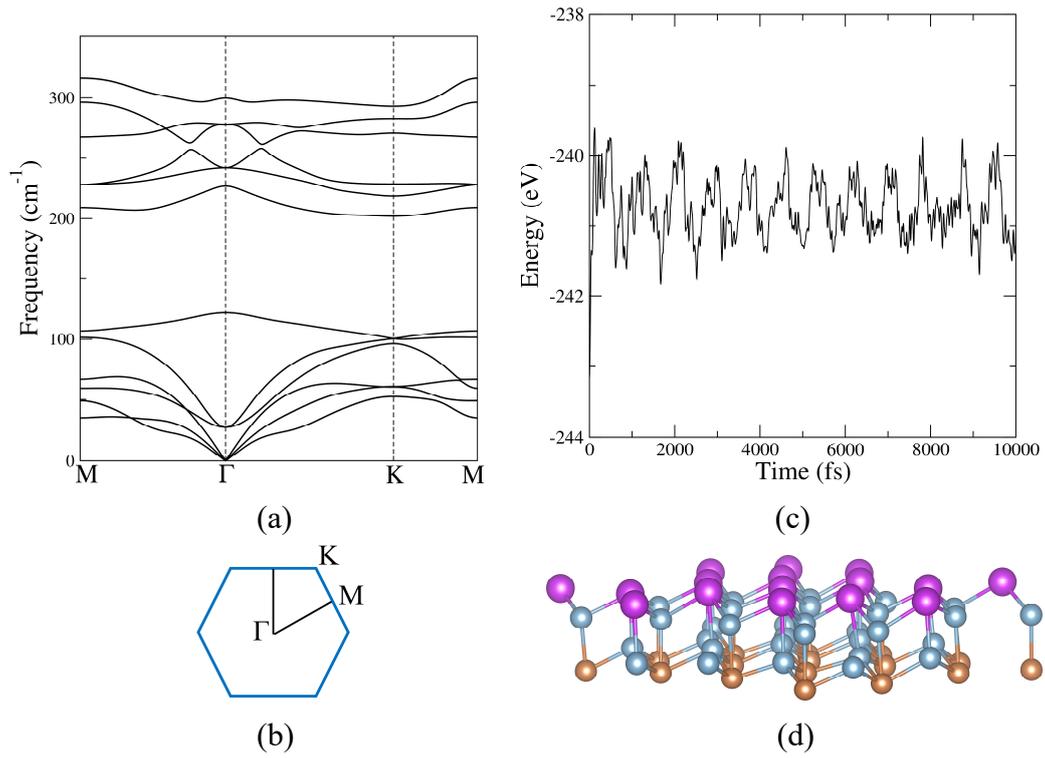

Fig. 2 Left panel: Phonon dispersion of monolayer Al$_2$SbBi (a) and high-symmetry points in the Brillouin zone of the hexagonal lattice (b). Right panel: Changes of energy with time obtained from the molecular dynamics simulation at 300 K in a canonical ensemble (c) and structure snapshot at the end of 10 ps (d).

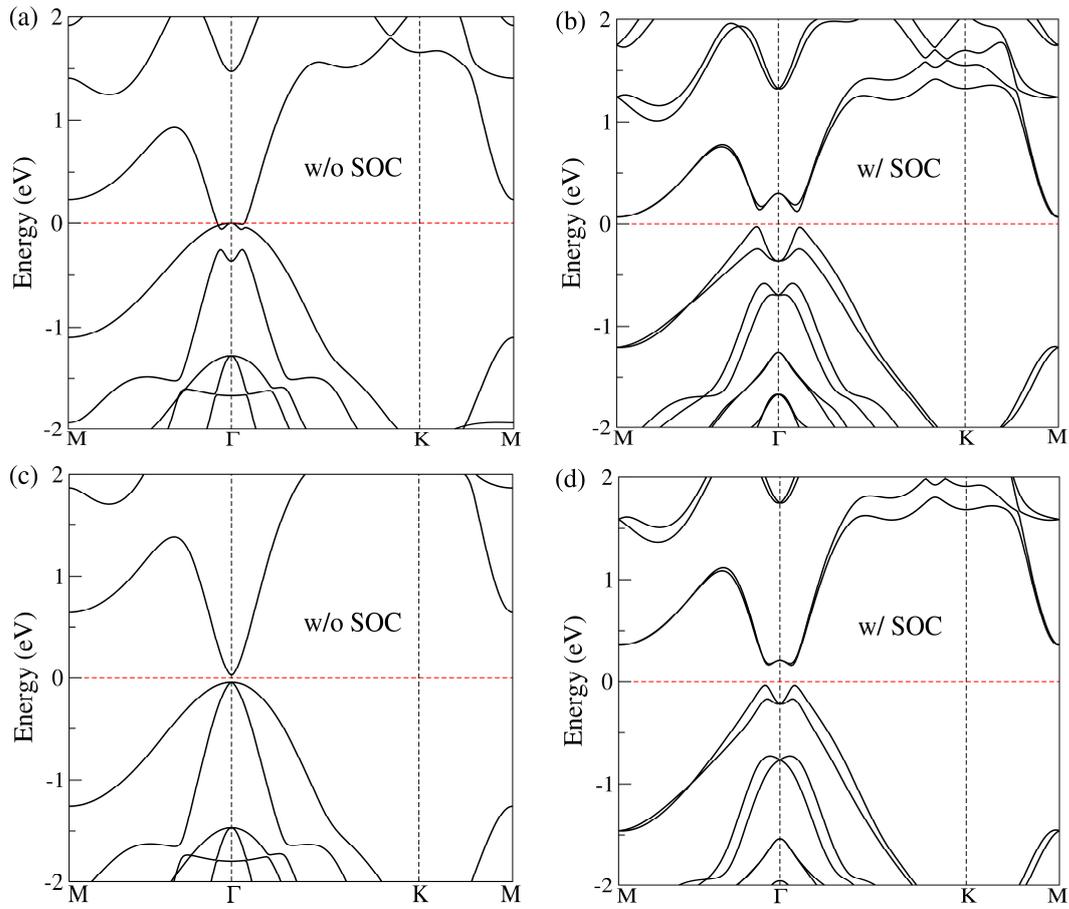

Fig. 3 Electronic band structure of monolayer Al$_2$SbBi without and with SOC calculated using the PBE functional [(a) and (b)] and HSE06 hybrid functional [(c) and (d)]. The fermi energy is set to zero.

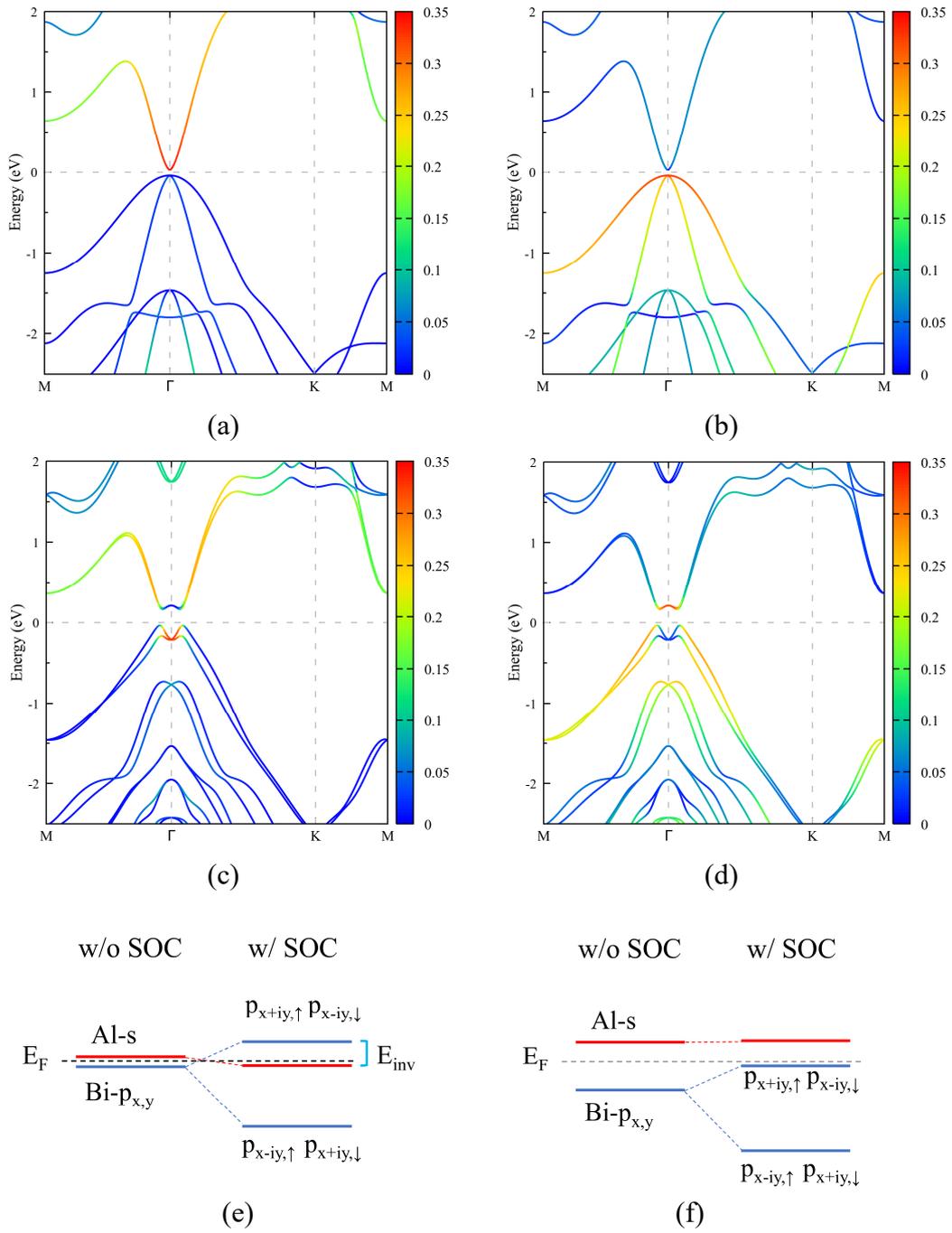

Fig. 4 Orbital decomposed HSE06 band structure without SOC: (a) Al:s and (b) Bi:$p_x+p_y$. Orbital decomposed HSE06 band structure including SOC: (c) Al:s and (d) Bi:$p_x+p_y$. The fermi energy is set to zero. Schematic diagram of the orbitals' evolution: (e) no strain and (f) compressive strain.

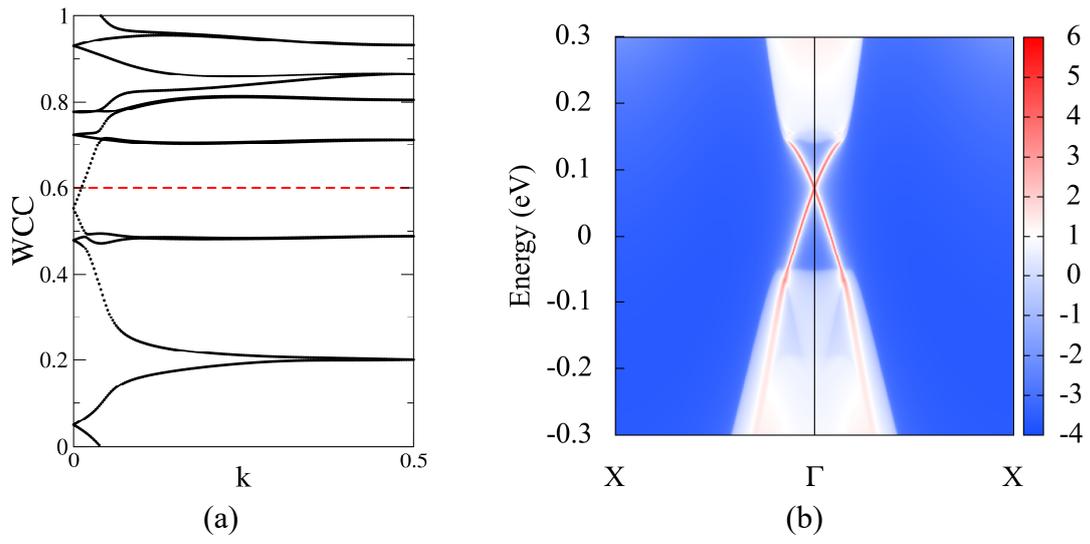

Fig. 5 Evolution of the WCC (a) and edge spectral density of the armchair edge (b) of monolayer $Al_2SbBi$.

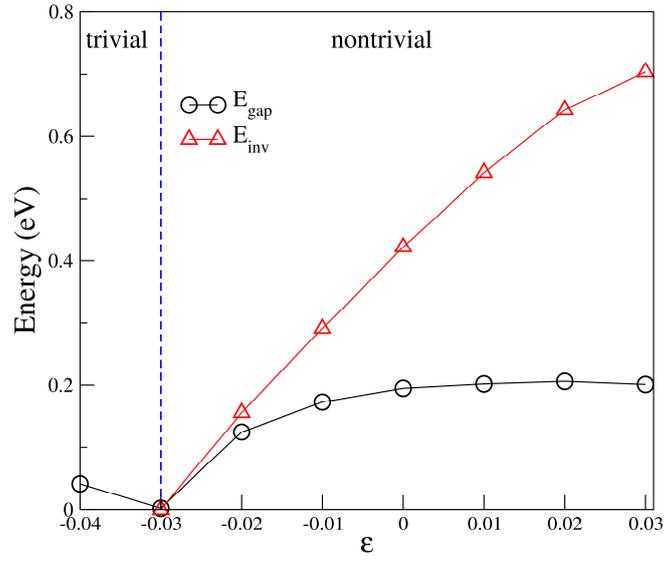

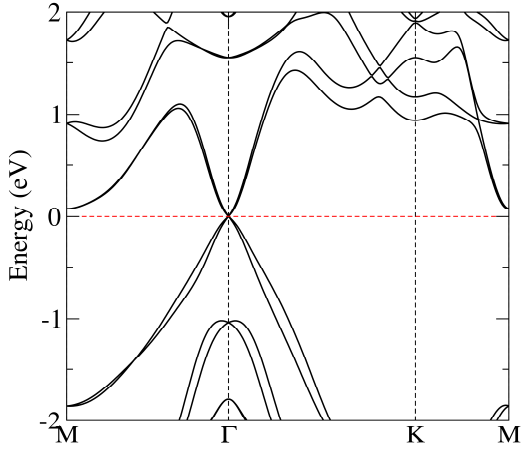  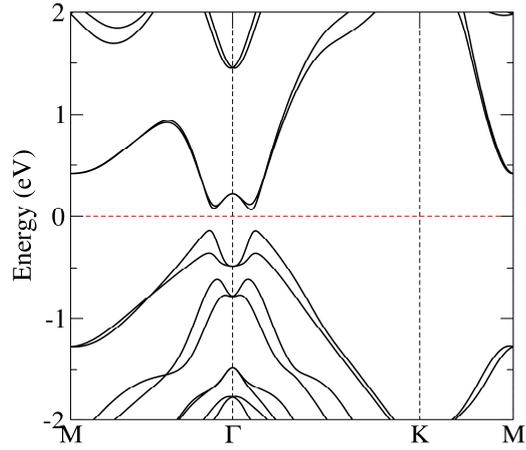

Fig. 6 The magnitude of the band inversion at the Γ point ($E_{inv}$) and the band gap ($E_{gap}$) as a function of biaxial strain ε (a), and the band structures for ε = -0.03 (b) and ε = 0.03 (c). All calculations are performed with the HSE06+SOC. The system turns into a trivial phase when ε ≤ -0.03.

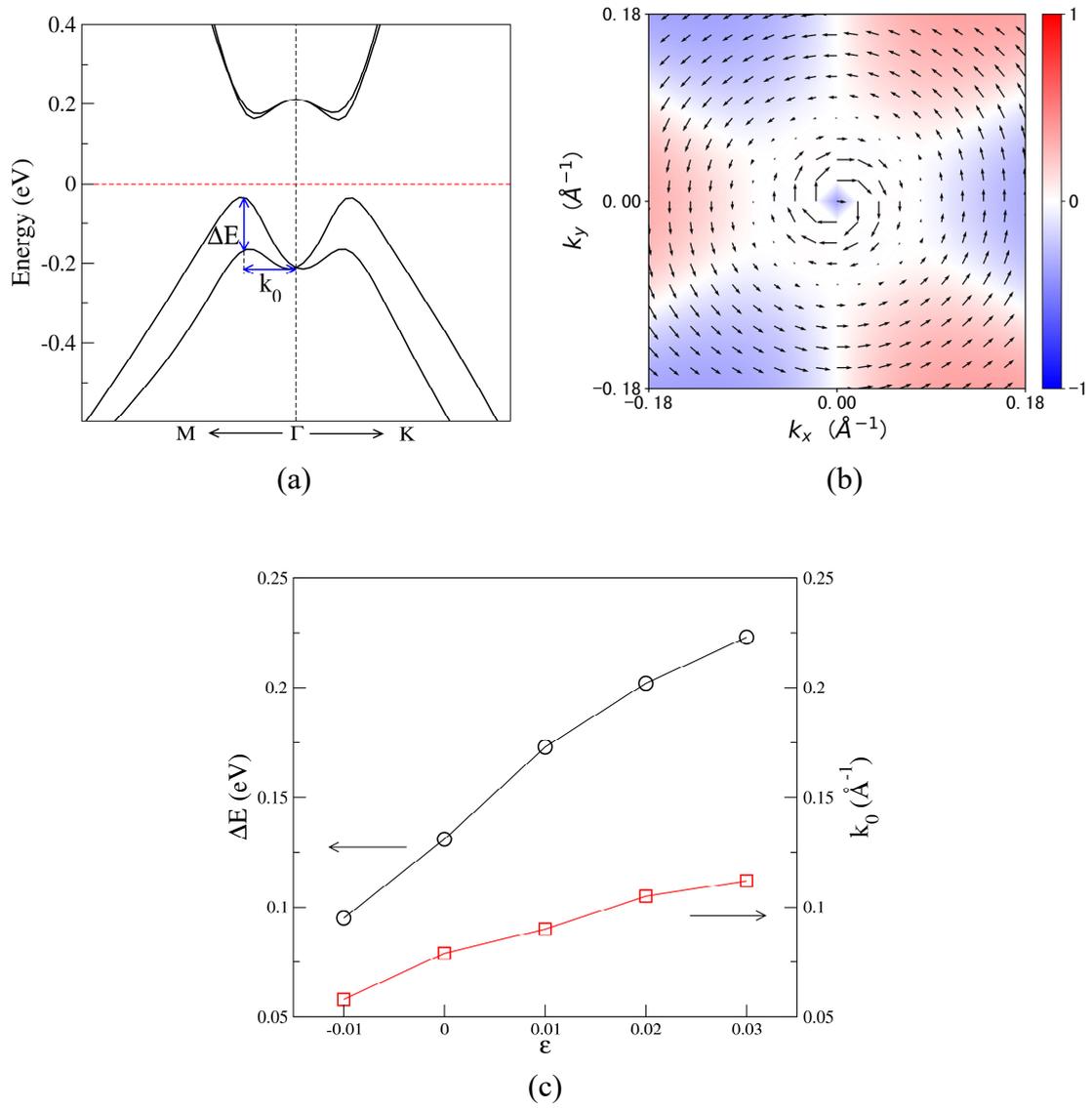

Fig. 7 Zoom in of the spin splitting at the top of valence bands (a) and the spin texture of the highest valence band around the Γ point (b). The arrows denote the in-plane spin direction and the color scheme indicates the out-of-plane spin component. (c) The magnitude of the spin splitting energy ΔE and momentum offset $k_0$ as a function of biaxial strain ε. All calculations are performed with the HSE06+SOC.

# Supplementary material:

# Large spin splitting and piezoelectricity in a two-dimensional topological insulator Al$_2$SbBi with double-layer honeycomb structure

D. Q. Fang[1*], H. Zhang[1], D. W. Wang[2,3]

[1]MOE Key Laboratory for Nonequilibrium Synthesis and Modulation of Condensed Matter, School of Physics, Xi'an Jiaotong University, Xi'an 710049, China

[2]School of Microelectronics & State Key Laboratory for Mechanical Behavior of Materials, Xi'an Jiaotong University, Xi'an 710049, China

[3]Key Lab of Micro-Nano Electronics and System Integration of Xi'an City, Xi'an Jiaotong University, Xi'an 710049, China

*fangdqphy@xjtu.edu.cn

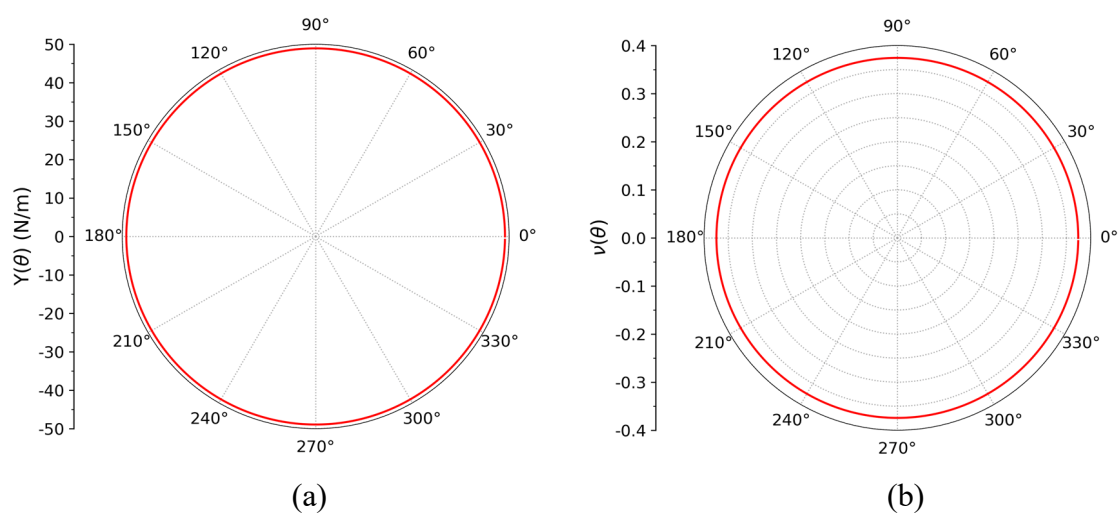

Fig. S1 Variation of the in-plane Young's modulus (a) and Poisson's ratio (b) with the crystal orientation of monolayer Al$_2$SbBi.

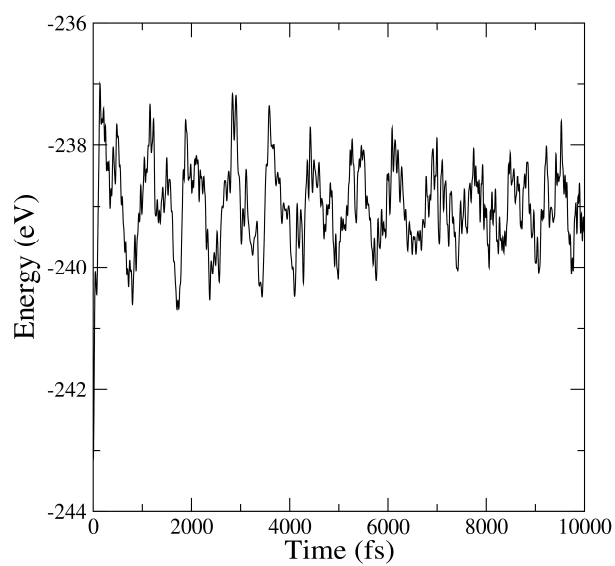

(a)

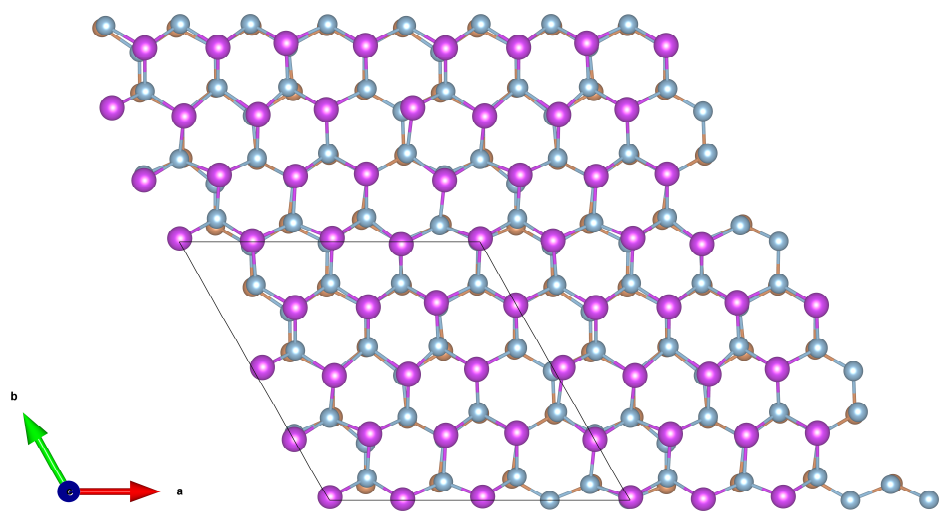

(b)

Fig. S2 (a) Changes of energy with time obtained from the molecular dynamics simulation at 500 K in a canonical ensemble and (b) structure snapshot at the end of 10 ps.

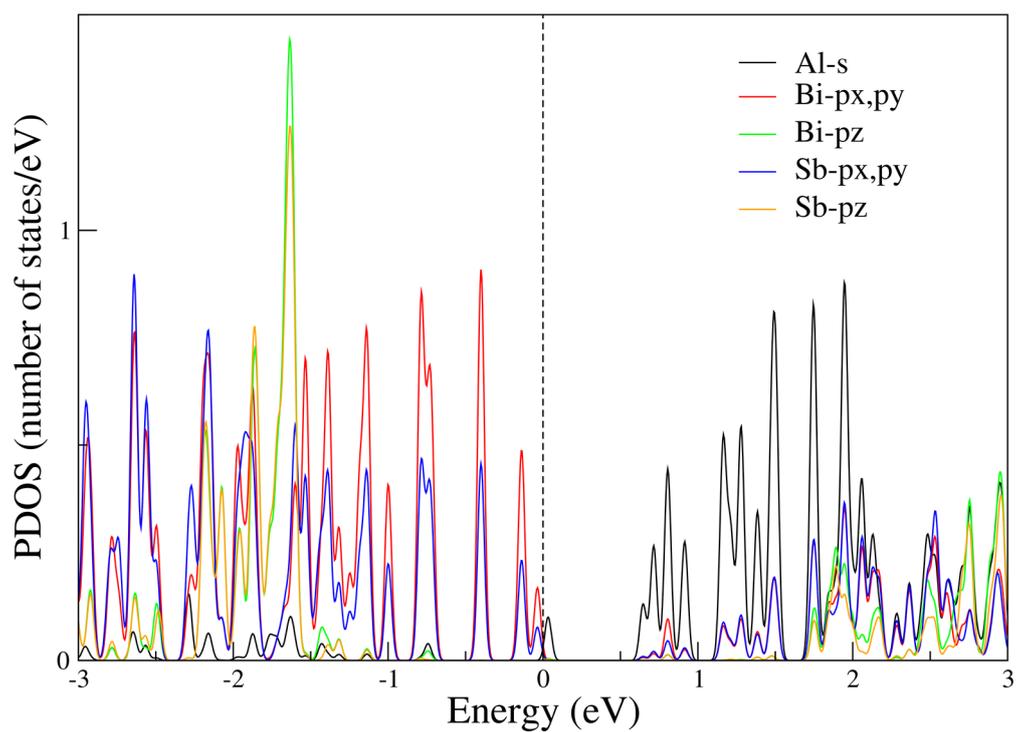

Fig. S3 Projected density of states (PDOS) of monolayer Al$_2$SbBi calculated using the HSE06 without SOC.

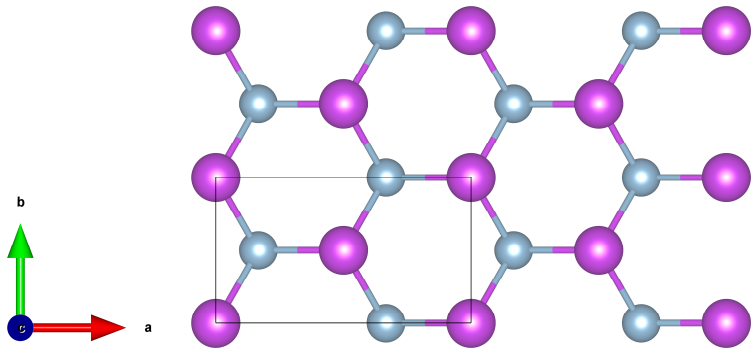

Fig. S4 The rectangular supercell of monolayer $Al_2SbBi$, indicated by the black frame, used for the piezoelectric coefficient calculations.